\documentclass[12pt]{article}
\usepackage{amsfonts}
\usepackage{amssymb}
\usepackage{amsbsy}
\usepackage{mathrsfs}
\usepackage{amsmath}
\usepackage{epsfig}      
\usepackage{color}       

\newlength{\TZ}
\newcommand{\BEQ}{\begin{equation}}     
\newcommand{\BEA}{\begin{eqnarray}}
\newcommand{\BD}{\begin{displaymath}}
\newcommand{\EEQ}{\end{equation}}       
\newcommand{\EEA}{\end{eqnarray}}
\newcommand{\ED}{\end{displaymath}}
\newcommand{\bb}{\begin{eqnarray}}
\newcommand{\ee}{\end{eqnarray}}

\newcommand{\vro}{\varrho}              
\newcommand{\D}{{\rm d}}                
\newcommand{\II}{{\rm i}}               

\newcommand{\demi}{\frac{1}{2}}         

\newcommand{\wht}[1]{\widehat{#1}}      
\newcommand{\lap}[1]{\overline{#1}}     

\renewcommand{\vec}[1]{\boldsymbol{#1}} 

\newcommand{\appsektion}[1]{\setcounter{equation}{0}\setcounter{subsection}{0}
\section*{Appendix. #1}
\renewcommand{\theequation}{A.\arabic{equation}}
              \renewcommand{\thesection}{A} }

\catcode`\@=11
\def\numberbysection{\@addtoreset{equation}{section}
        \def\theequation{\thesection.\arabic{equation}}}
\numberbysection

\definecolor{gruen}{rgb}{0,0.625,0}       
\definecolor{rot}{rgb}{0.75,0,0}          
\definecolor{blau}{rgb}{0,0,0.75}         
\definecolor{casta}{rgb}{0.45,0.20,0}     
\definecolor{gelb}{rgb}{0.825,0.725,0.0}  

\begin{document}

\begin{titlepage}

\vskip 1.5 cm
\begin{center}
{\Large \bf Non-equilibrium relaxations:\\[0.1cm] ageing and finite-size effects}\footnote{Dedi\'e \`a Bertrand Berche pour son 60$^e$ anniversaire.} 
\end{center}

\vskip 2.0 cm
\centerline{ {\bf Malte Henkel}$^{a,b}$
}
\vskip 0.5 cm
\begin{center}
$^a$Laboratoire de Physique et Chimie Th\'eoriques (CNRS UMR 7019),\\  Universit\'e de Lorraine Nancy,
B.P. 70239, F -- 54506 Vand{\oe}uvre l\`es Nancy Cedex, France\\~\\
$^b$Centro de F\'{i}sica Te\'{o}rica e Computacional, Universidade de Lisboa, \\Campo Grande, P--1749-016 Lisboa, Portugal\\~\\
\end{center}

\begin{abstract}
The long-time behaviour of spin-spin correlators in the slow relaxation of systems undergoing phase-ordering kinetics 
is studied in geometries of finite size. A phenomenological
finite-size scaling ansatz is formulated and tested through the exact solution of the kinetic spherical model, 
quenched to below the critical temperature, in $2<d<4$ dimensions.
~\\~\\
\end{abstract}

\vfill
PACS numbers: 05.40.-a, 05.70.Ln, 05.70.Jk, 75.40.Gb  

\end{titlepage}

\setcounter{footnote}{0}

\section{Critical relaxations in finite-size systems}

Collective phenomena arise in many-body systems with dynamically created long-range interactions and thereby often show new 
qualitative properties which cannot be obtained in systems with a small number of degrees of freedom. An important class are 
{\em critical phenomena}, characterised by {\em scale-invariance}. We are interested here in time-dependent phenomena with
time-dependent or `dynamical' scaling. As a physical example, we consider many-body spin systems, initially prepared in a 
disordered state with at most short-ranged correlations 
and then suddenly quenched to a temperature $0<T<T_c$ below the critical temperature $T_c>0$, with at least two physically 
distinct phases. Such a quenched spin system is then said to undergo {\em phase-ordering kinetics} \cite{Bray94a}. For a 
spatially infinite geometry, observables such as correlation functions are then expected to be invariant under the time-space 
dilatation
\BEQ \label{1.0}
t \mapsto t' = \kappa^z t \;\; , \;\; \vec{r} \mapsto \vec{r}' = \kappa \vec{r}
\EEQ
where $\kappa$ is a constant re-scaling factor and the {\em dynamical exponent} $z$ serves to distinguish the scaling between 
time and space. The relaxation of the system after the quench can be measured through connected correlators of the time- 
and space-dependent spin variables $S_{\vec{r}}(t)$, namely  
\begin{subequations} \label{1.1}
\begin{align} 
 C(t;\vec{r}) &:= \left\langle S_{\vec{r}}(t)S_{\vec{0}}(t)\right\rangle 
                 - \left\langle S_{\vec{r}}(t)\right\rangle\left\langle S_{\vec{0}}(t)\right\rangle 
 = F_C\left(\frac{|\vec{r}|}{t^{1/z}}\right) \label{1.1a} \\
 C(t,s) &:= \left\langle S_{\vec{r}}(t)S_{\vec{r}}(s)\right\rangle 
           - \left\langle S_{\vec{r}}(t)\right\rangle\left\langle S_{\vec{r}}(s)\right\rangle 
 = f_C\left(\frac{t}{s}\right) \label{1.1b}
\end{align}
\end{subequations}
where the quoted scaling forms are meant to hold in the limit of large times and large distances, such that $|\vec{r}|^z/t$ 
and $t/s$ are kept fixed. In (\ref{1.1b}), $t$ is the {\em observation time} and $s$ is the {\em waiting time}. 
Asymptotically, the scaling function $f_{C}(y)$ in (\ref{1.1b}) should be algebraic
\BEQ \label{1.2}
f_C(y) \sim y^{-\lambda/z} \;\; , \;\; \mbox{\rm as $y\to\infty$}
\EEQ
where $\lambda=\lambda_C$ is the {\em autocorrelation exponent}. 
A many-body non-stationary system whose slow relaxation dynamics also breaks time-translation-invariance and is such that the 
single-time correlator $C(t,\vec{r})$ and the two-time auto-correlator $C(t,s)$ obey the dynamical scaling (\ref{1.1}), 
is said to be {\em ageing} \cite{Stru78,Cugl03,Henk10,Taeu14}. 

\begin{table}[tb]
\begin{center}
\begin{tabular}{|llr|ll|l|} \hline
\multicolumn{3}{|l|}{~material/model}      & $\,z$~     & ~~$\lambda$~ & \multicolumn{1}{c|}{Refs.} \\  \hline
\multicolumn{2}{|l}{Merck (CCH-501)} &     & $1.94(5)$  & $1.246(79)$ & ~\cite{Maso93} \\
\multicolumn{2}{|l}{nematic TNLC}    &     & $2.01(1)$  & $1.28(11)$  & ~\cite{Alme21} \\ \hline 
~Ising   &~$1D$     & {\sc lr}             & $1+\sigma$ & $0.5$       & ~\cite{Corb19a,Corb19b}\\[0.12cm]
~Ising   &~$2D$     & {\sc lr}             & $1+\sigma$ & $1$         & ~\cite{Chris19,Chris20} \\[0.12cm]
~Ising   &~$2D$     & {\sc sr}             & 2          & $1.24(2)$   & ~\cite{Lorenz07a} \\
         &          &                      & 2          & $1.25$      & ~\cite{Chris19,Chris20} \\
         &          &                      & 2          & $1.3$       & ~\cite{Maze90,Midya14,Vada19}\\[0.12cm]
~Ising   &~$3D$     & {\sc sr}             & $2$        & $1.60(2)$   & ~\cite{Henk03}\\
         &          &                      & $2$        & $1.6$       & ~\cite{Maze90,Midya14}\\[0.12cm]
~Potts-3 &~$2D$     & {\sc sr}             & $2$        & $1.19(3)$   & ~\cite{Lorenz07a} \\
         &          &                      & $2$        & $1.22(2)$   & ~\cite{Corb06a} \\[0.12cm]
~Potts-8 &~$2D$     & {\sc sr}             & $2$        & $1.25(1)$   & ~\cite{Lorenz07a}\\[0.12cm]
~$XY$    & ~$3D$    & {\sc sr}             & $2$        & $1.7(1)$    & ~\cite{Abri04}\\
         &          &                      & $2$        & $1.6$       & ~\cite{Maze90} \\[0.12cm]
\multicolumn{2}{|l}{~spherical} & {\sc sr} & $2$        & $d/2$       & ~\cite{Godr00b} \\
\multicolumn{2}{|l}{~spherical} & {\sc lr} & $\sigma$   & $d/2$       & ~\cite{Cann01,Baum07} \\ \hline
\end{tabular} \end{center}
\caption[Autocorrelation exponent]{\small Dynamical exponent $z$ and autocorrelation exponent $\lambda$, as measured 
experimentally or found in some spin models. 
Long-range ({\sc lr}) behaviour occurs in the Ising model for $\sigma<1$ and in the spherical model for $\sigma<2$. 
The spherical model is considered for dimensions $d>z$. 
\label{Tab:lambda0}
}
\end{table}

For phase-ordering, with a non-conserved order parameter, some general exact results exist for models with short-ranged interactions. 
First, the dynamical exponent $z=2$ for a non-conserved order parameter \cite{Bray94b}.\footnote{In this work, we restrict to this {\em model-A dynamics}.}   
Second, the Yeung-Rao-Desai inequality states that $\lambda\geq d/2$ \cite{Yeun96a}. 
Third, for the $2D$ Ising model one has the Fisher-Huse inequality $\lambda\leq \frac{5}{4}$ \cite{Fish88a}. 
Some typical values for $z$ and $\lambda$ are listed in table~\ref{Tab:lambda0}. 
They illustrate the sharpness of these exact bounds and permit a comparison between
short-ranged and long-ranged interactions. The agreement with the available experiments \cite{Maso93,Alme21} is very satisfying. 
For more detailed tables, see \cite{Henk10}. 

How is the scaling behaviour, encoded in the scaling forms (\ref{1.1}), modified in a system confined to a domain of finite size, 
e.g. because it is placed into a box~?

\begin{figure}[tb]
\begin{center}
\includegraphics[width=.46\hsize]{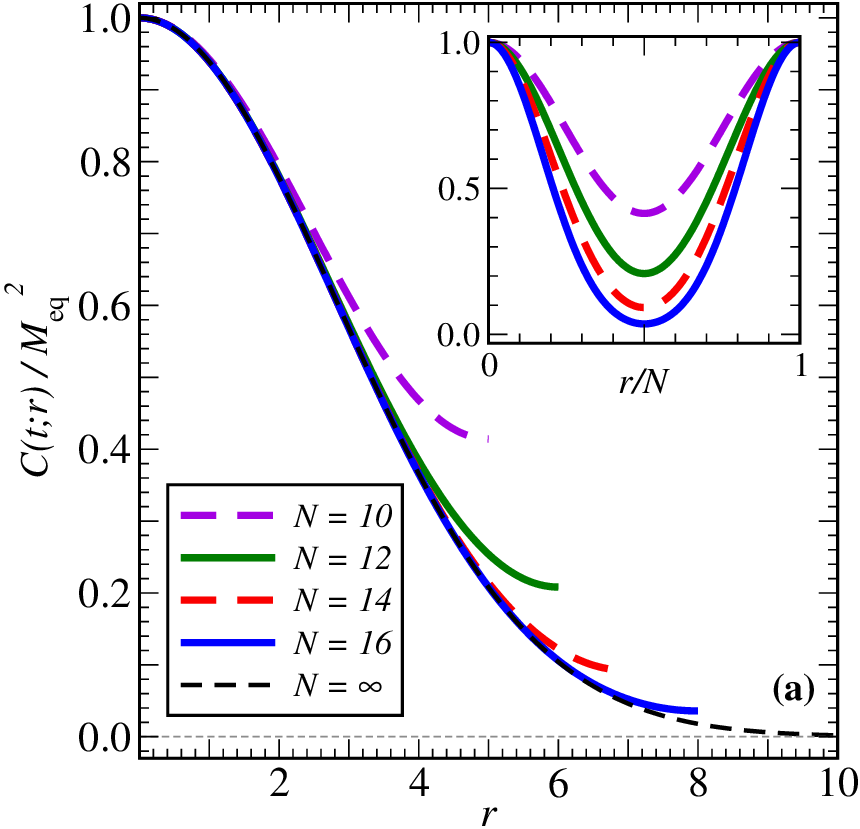} ~\includegraphics[width=.46\hsize]{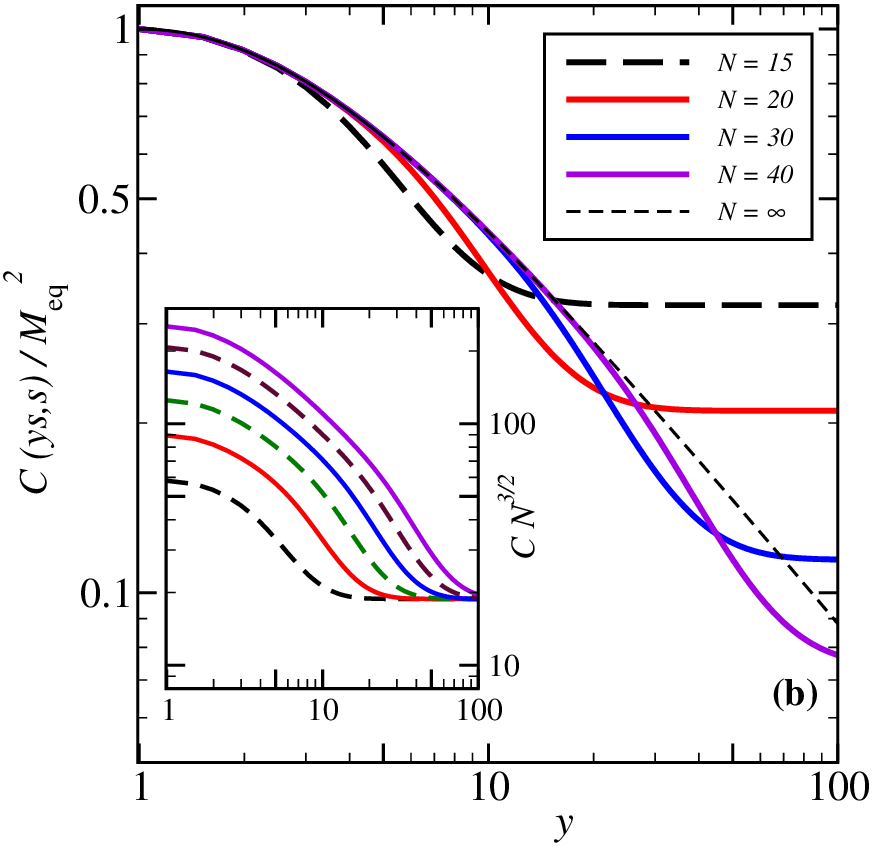}
\end{center}
\caption[fig1B]{\small{\bf (a)} Finite-size effects for the single-time correlator $C(t,\vec{r})$ in the fully finite spherical model at $T<T_c$, 
with $t=50$ and $N=[10,12,14,16,\infty]$ from top to bottom. The inset shows the periodicity over the interval $0\leq r/N\leq 1$. \\
{\bf (b)} Finite-size effects for the two-time autocorrelator $C(ys,s)$ in the $3D$ fully finite spherical model at $T<T_c$, for
$N=[15,20,30,40]$ from top to bottom (at the right) and $s$ fixed. The thin dashed line gives the infinite-size autocorrelator. The inset shows
the data collapse of the re-scaled correlator $C N^{3/2}$ for $y=t/s$ large, with $N=[15,20,25,30,35,40]$ from left to right (arbitrary units). 
\label{fig1B} }
\end{figure}

For a phenomenological answer, consider figure~\ref{fig1B}. For a fully finite hyper-cubic lattice with $N^d$ sites and periodic boundary conditions, 
the single-time correlator $C(t;\vec{r})$ is shown in figure~\ref{fig1B}a, where $\vec{r}$ is oriented along one of  coordinate axes. 
If the spatial distances $r=|\vec{r}|$ are not too large, the shape of the correlator does not depend sensitively on $N$. 
Only if $r\lesssim \frac{N}{2}$, does the correlator also receive contributions `from around the world', such that for $r\approx \frac{N}{2}$ 
it no longer tends towards zero, but rather saturates at a $N$-dependent constant $C^{(1)}_{\rm lim}(N)>0$. Figure~\ref{fig1B}b shows the
two-time autocorrelator $C(ys,s)$. For large $s$, but $y$ small enough, there is a clear data collapse. 
However, for larger values of $y$, $C$ begins to decrease more
rapidly than the infinite-size curve (\ref{1.1b}).\footnote{Since for lattices large enough that the system is just leaving the effective finite-size
regime, the local exponent estimates $\lambda_{\rm eff}(y)$ may slightly over-estimate $\lambda$. In certain cases this might lead to
claims of violation of exact upper bounds such as the Fisher-Huse inequality.} 
As $y\gg 1$, $C$ finally saturates at the limit value $C^{(2)}_{\infty}(N)>0$. 
 
\begin{figure}[tb]
\begin{center}
\includegraphics[width=.48\hsize]{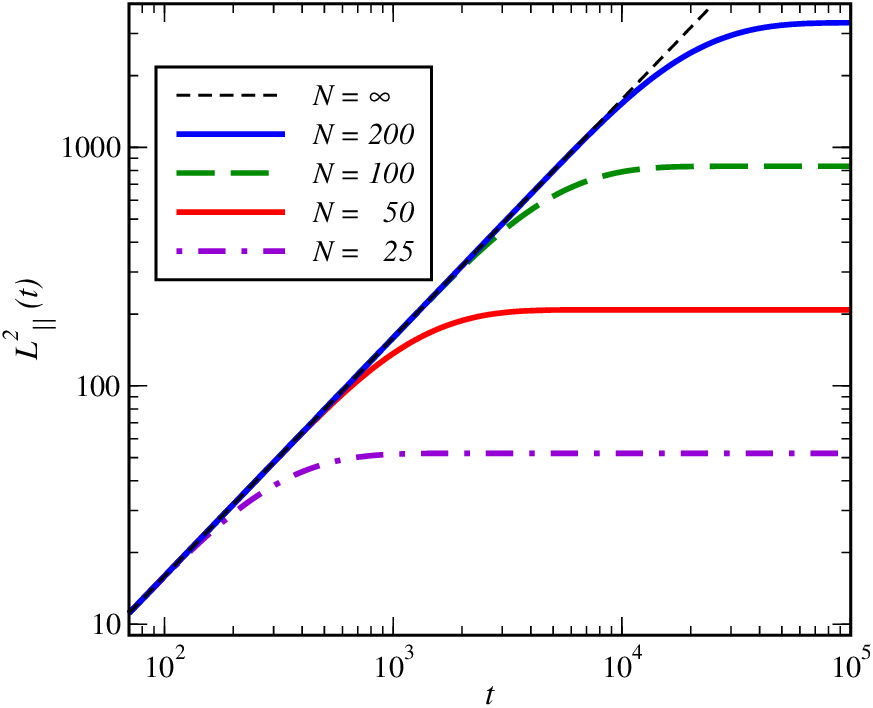}
\end{center}
\caption[fig2B]{\small Finite-size effects for the longitudinal characteristic length $L_{\|}^2(t)$, 
measured along a coordinate axis, in the fully finite spherical model with 
lattice sizes $N=[25,50,100,200]$ from bottom to top. The thin dashed line indicates the infinite-size behaviour $L(t)\sim t^{1/2}$.  
\label{fig2B} }
\end{figure}

Although the single-time correlator does not display strong finite-size effects, this is different for the length scale $L=L(t)$ of the
growing clusters, estimated from the second moment 
\BEQ \label{1.4}
L^2(t) = \frac{\sum_{\vec{r}} |\vec{r}|^2 C(t;\vec{r})}{\sum_{\vec{r}}  C(t;\vec{r})} 
\EEQ
The precise extent of the sums will be specified below. Figure~\ref{fig2B} shows that for sufficiently short times, 
the length $L^2(t)\sim t$ behaves as for the
infinite system, but as $t$ grows further, finally there occurs a cross-over towards a finite constant $L_{\infty}(N)$. 
We shall see how to explain the findings of figures~\ref{fig1B} and~\ref{fig2B} in terms of phenomenological finite-size scaling. 
The resulting predictions will be tested in the exactly solved kinetic spherical model, for dimensions $2<d<4$.

The {\em spherical model} of a ferromagnet \cite{Berl52,Lewi52} has served as an exactly solvable, yet non-trivial,
model for the detailed analysis of general concepts of critical phenomena, see \cite{Fish05} for a historical perspective. 
Its non-equilibrium behaviour after a quench has also been thoroughly analysed, see 
\cite{Ronc78,Cugl94,Cugl95,Godr00b,Cann01,Fusc02,Pico02,Hase06,Anni06,Baum07,Dutt08,Anni09,Henk10}. 
The related Arcetri model provides a qualitative description of the dynamics in the non-equilibrium growth
of interfaces \cite{Henk15,Dura17}. Finite-size effects at equilibrium have also been analysed at great depth 
in the spherical model and have been of value to test the theory of finite-size scaling derived from the renormalisation group, 
see \cite{Barb73,Brez82,Barb83,Luck85,Singh85,Singh87,Priv90,Alle93,Bran00,Cham08} and refs. therein. 
For dimensions $d>d_c=4$, that is above the upper critical dimension, the
standard finite-size scaling {\em ansatz} must be considerably modified \cite{Bind85,Priv90,Kenn14,Flor15,Flor16,Grim17,Berc22}.

Finite-size scaling techniques have been applied in studies of
phase-ordering kinetics \cite{Midya14,Vada19}, the ageing of polymer collapse \cite{Maju16,Chris17,Maju17,Maju20} 
or the dynamics of mitochondrial networks \cite{Zamp22}. Explicit studies of finite-size scaling in an ageing system 
have been carried out in Ising spin glasses \cite{Komori99} and notably 
on the dimensional cross-over between the 
$3D$ and $2D$ Edwards-Anderson spin glass \cite{Fern19} as motivated by extremely
accurate experiments on CuMn films \cite{Zhai17,Zhai19}. In addition,  
finite-size effects analogous to figures~\ref{fig1B} and~\ref{fig2B} 
are clearly visible in the time-evolution of characteristic cluster sizes in long-ranged Ising models
quenched to $T<T_c$ \cite{Chris19} or in the auto-correlator \cite{Chris20}. 
Since the bulk $3D$ spherical model and the bulk $(p=2)$ spherical
spherical spin glass are in the same dynamic universality class \cite{Cugl95}, one might hope that finite-size effects could be similar as well. Not so~! 
Rather, detailed studies of the $(p=2)$ spherical spin glass \cite{Fydo15,Barbier21} 
show that this equivalence only holds in the spin glass for times $t\ll t_{\rm cross} \sim N^{2/3}$. 
For time scales $t\gg t_{\rm cross}$, ageing still holds with a new set of universal exponents \cite{Fydo15}, 
to be followed by a second cross-over to a regime of exponential decay at extremely large times \cite{Barbier21}.

This work is organised as follows. In section~2, we recall the main features of dynamical scaling in ageing phase-ordering kinetics. 
In section~3, we extend this phenomenological treatment to finite systems, using the hyper-cubic geometry 
$\overbrace{N\times\cdots \times N\,}^{d^*\: \mbox{\rm\scriptsize factors}}\times 
\overbrace{\infty \times \cdots \times \infty}^{d-d^*\: \mbox{\rm\scriptsize factors}}$, 
where the first $d^*\leq d$ directions are finite and periodic and the other $d-d^*$ directions are infinite. 
The finite-size forms so obtained will be checked in section~4 using the exact solution of the kinetic spherical model in $2<d<4$ dimensions, 
quenched to $T>T_c$ from a totally disordered state and in section~5 we conclude. Technical details of the exact solution are given in
the appendix. 

\section{Dynamical scaling description}

A central ingredient of ageing is dynamical scaling. For the general two-time and spatial bulk correlator, our starting point is (below the upper
critical dimension $d<d_c$; for short-ranged interactions usually $d_c^{({\rm short})}=4$)
\BEQ \label{2.1}
C\left(\kappa^z t, \kappa^z s; \kappa \vec{r}\right) = \kappa^{\phi} C\left(t,s;\vec{r}\right)
\EEQ
where $t,s$ are the observation and the waiting time, $z$ is the dynamical exponent, $\phi$ a scaling exponent and 
$\vec{r}$ is the spatial distance. Writing (\ref{2.1}) means that we assume negligible all finite-time and finite-distance corrections to scaling. 
Choosing $\kappa=s^{-1/z}$, this gives
\BEQ \label{2.2}
C\left(t,s;\vec{r}\right) = s^{\phi/z} C\left(\frac{t}{s}, 1;\frac{\vec{r}}{s^{1/z}}\right)
\EEQ
In phase-ordering, the single-time correlator at $\vec{r}=\vec{0}$ is finite; 
namely either $C(t;\vec{0})=1$ in Ising-like systems or else $C(t;\vec{0})=M_{\rm eq}^2$ for order
parameters with a continuous global symmetry. Setting $s=t$ in (\ref{2.2}), this leads to $\phi=0$ and\footnote{If more generally, 
one would expect $C(t,s) = s^{-b} f_C(t/s)$, this would lead to the identification $b=-\phi/z$, but for $\phi\ne 0$, this is incompatible with
$C(t;\vec{0})$ being finite and constant for $t\to\infty$.} further to 
$C(t;\vec{r})=C\bigl(1,1;|\vec{r}| t^{-/z}\bigr)=:F_C(|\vec{r}| t^{-/z}\bigr)$. On the other hand, setting now $\vec{r}=\vec{0}$, 
the two-time auto-correlator is $C(t,s) = C\bigl(t,s;\vec{0}\bigr)=C(t/s,1;\vec{0}) =: f_C(t/s)$. These results fully reproduce (\ref{1.1}). 

\section{Dynamical finite-size scaling}

According to the original definition, {\em finite-size scaling} \cite{Fish71} is the scaling behaviour in a nearly critical system confined to a
{\em geometry of finite linear extent}\, $N$. For finite geometries, the natural generalisation of (\ref{2.1}) consists, as at equilibrium 
\cite{Barb73,Suzuki77,Barb83,Priv90}, to consider $1/N$ as a further relevant scaling field.\footnote{Very interesting adaptations of this idea have been 
brought forward in the study of the kinetics of polymer collapse, where $N$ is now the finite number of monomers, 
but the spatial geometry of the system was not specified \cite{Maju17,Maju20}.}  
While this hypothesis was originally specified for the order parameter at the critical point \cite{Suzuki77}, 
we adapt this to the situation at hand and write down the finite-size scaling 
({\sc fss}) {\em ansatz} for the full two-time correlator
\BEQ \label{3.1}
C\left(\kappa^z t, \kappa^z s; \kappa \vec{r};\kappa^{-1} \frac{1}{N}\right) = \kappa^{\phi} C\left(t,s;\vec{r};\frac{1}{N}\right)
\EEQ
meant to hold in the hyper-cubic geometry $\overbrace{N\times\cdots \times N\,}^{d^*\: \mbox{\rm\scriptsize factors}}\times 
\overbrace{\infty \times \cdots \times \infty}^{d-d^*\: \mbox{\rm\scriptsize factors}}$
where $N$ describes the finite length in the system. For simplicity, we consider a single length of this kind.\footnote{Spatially 
anisotropic finite-size effects could be taken into account by introducing distinct finite sizes $N_j$ in different spatial directions.} 
Of course, for $N\to\infty$, one is back to the bulk scaling form (\ref{2.1}), and hence (\ref{1.1}). 

Choose the re-scaling factor $\kappa=s^{-1/z}$. For phase-ordering kinetics, recall from section~2 that $\phi=0$. 
Then (\ref{3.1}) can be equivalently expressed as
\BEQ \label{3.1bis}
C\left(t,s;\vec{r};\frac{1}{N}\right) = 
C\left(\frac{t}{s},1;\frac{\vec{r}}{s^{1/z}}; \frac{s^{1/z}}{N}\right)
\EEQ
As above in section~2, we then expect for the correlators (provided spatial rotation-invariance can be assumed) 
\BEQ \label{3.2}
C\bigl(t;\vec{r};N^{-1}\bigr) = F_C\left( \frac{|\vec{r}|^z}{t};\frac{N^z}{t}\right)
\;\; , \;\; 
C\bigl(t,s;N^{-1}\bigr) = f_C\left( \frac{t}{s};\frac{N^z}{t}\right) 
\EEQ
such that the corresponding scaling functions are now functions of two variables. Finite-size scaling in ageing can be analysed in the asymptotic
{\sc fss} limit where $t\to\infty$, $s\to \infty$, $|\vec{r}|\to\infty$ and $N\to\infty$ such that the three scaling variables
\BEQ \label{3.3}
y = \frac{t}{s} \;\; ~,~ \;\;
\vec{\vro} =  \frac{\vec{r}}{t^{1/z}} \;\; ~,~ \;\;
Z = \frac{N^z}{t}
\EEQ 
are kept fixed. The precise form of the finite-size scaling functions (\ref{3.2}) will depend on the universality class under study,
and on the boundary conditions \cite{Barb83,Priv90,Bran00}. 

As a first consequence, consider the characteristic length $L(t)$ of the clusters. From (\ref{1.4}) and (\ref{3.2}), we derive the finite-size scaling form
\BEQ \label{3.4}
L^2(t;N^{-1}) = \frac{\sum_{\vec{r}} |\vec{r}|^2 C(t;\vec{r};N^{-1})}{\sum_{\vec{r}}  C(t;\vec{r};N^{-1})} 
\simeq t^{2/z} 
\frac{\int\!\D\vec{r}\: \bigl( |\vec{r}| t^{-1/z}\bigr)^{2} F_C(|\vec{r}|^z/t;N^z/t)}{\int\!\D\vec{r}\: F_C(|\vec{r}|^z/t;N^z/t)} 
= t^{2/z} f_{L}\left(\frac{N^z}{t}\right)
\EEQ
For $Z\gg 1$, the behaviour of an effectively infinite system requires that $f_L(Z) \stackrel{Z\gg 1}{\simeq} f_0 = \mbox{\rm cste.}$ and for
$Z\ll 1$, the time-independent saturation in figure~\ref{fig2B} is captured by $f_L(Z)\stackrel{Z\ll 1}{\sim} Z^{2/z}$ such that $L_{\infty}(N)\sim N$, 
as would have been expected from dimensional analysis.

Next, we consider the plateau in the two-time auto-correlator $C(ys,s)$ for $y\gg 1$, see figure~\ref{fig1B}b. 
Recall that for the infinite system, we expect from (\ref{1.1b},\ref{1.2}) that $C(t,s;\vec{0};0)=f_C(t/s)\sim \bigl(t/s\bigr)^{-\lambda/z}$. 
For $N<\infty$, we reformulate (\ref{3.1bis}) as follows  
\BEQ
C\bigl( t,s;N^{-1}\bigr) = C\left(t,s;\vec{0};\frac{1}{N}\right) = C\left(\frac{t}{s},1;\vec{0}; \frac{s^{1/z}}{N}\right) 
= \left(\frac{t}{s}\right)^{-\lambda/z} \mathscr{F}_C\left( \left(\frac{t}{s}\right)^{1/z}, \frac{s^{1/z}}{N} \right)
\EEQ
Herein, the first argument in the scaling function $\mathscr{F}_C=\mathscr{F}_C\bigl(y,u\bigr)$ will be considered large and be kept fixed, $y\gg 1$. 
In that case, the scaling function will describe the cross-over between (i) the infinite-system behaviour 
(when $u=s^{1/z}/N\to 0$) $f_C(y) = \mathscr{F}_C(y,0)\sim y^{-\lambda/z}$  which is independent of $s$ and 
(ii) the fully finite-system behaviour (when $u=s^{1/z}/N\to\infty$) when $C\stackrel{y\gg 1}{\longrightarrow} C^{(2)}_{\infty}$ 
no longer depends on $y=t/s$. 
The first limit case is taken into account by admitting $\mathscr{F}_C(y,u)\simeq \mathtt{F}\bigl(y u\bigr)$ and $\mathtt{F}(0)=\mbox{\rm cste}$. 
Then the second limit case leads to 
\BEQ
C\left(t,s;\vec{0};\frac{1}{N}\right) 
\stackrel{t/s\gg 1}{\simeq}\left(\frac{t}{s}\right)^{-\lambda/z}\mathtt{F}\left( \left(\frac{t}{s}\right)^{1/z}\cdot \frac{s^{1/z}}{N} \right)
\sim \left(\frac{t}{s}\right)^{-\lambda/z} \left( \left(\frac{t}{s}\right)^{1/z} \cdot\frac{s^{1/z}}{N} \right)^{\omega}
\EEQ
where in the last step, we assumed a power-law form of $\mathtt{F}(yu)\sim \bigl(y u\bigr)^{\omega}$ for $yu\gg 1$. 
The $y$-independent plateau $C^{(2)}_{\infty}$ observed for fully finite systems 
(see figure~\ref{fig1B}b for $s$ fixed) is reproduced if we choose $\omega=\lambda$. Hence, for finite systems with $y=t/s\gg 1$
\BEQ \label{3.11}
C\left(t,s;\vec{0};\frac{1}{N}\right) \stackrel{t/s\gg 1}{\longrightarrow} C^{(2)}_{\infty} \sim \left( \frac{s^{1/z}}{N}\right)^{\lambda}
\EEQ
Herein, $s$ is still kept fixed whereas $N$ must be taken large enough such that the system under study is indeed 
in its finite-size scaling regime (in other word, $N s^{-1/z}$ must be large enough). 

Hence for fully finite systems, quenched to $T<T_c$, the auto-correlator $C(ys,s)=f_C(y)\stackrel{y\gg 1}{\longrightarrow} C^{(2)}_{\infty}$, 
such that the plateau value $C^{(2)}_{\infty}=C^{(2)}_{\infty}(s,N)$ should obey the scalings
\BEQ \label{3.12}
C^{(2)}_{\infty} \sim N^{-\lambda} \mbox{\rm ~~with $s$ fixed}\;\; ~~,~~ \;\; 
C^{(2)}_{\infty} \sim s^{\lambda/z} \mbox{\rm ~~ with $N$ fixed}   
\EEQ
These are the sought scalings for the plateau of the autocorrelator and the main result of this section. 

The inset in figure~\ref{fig1B}b shows the data collapse of $N^{\lambda}C(ys,s)$ to a $y$-independent constant for $y$ large enough and $s$ fixed, in the
$3D$ spherical model, where $\lambda=\frac{3}{2}$. 
In the next section, (\ref{3.12}) will be verified analytically from the exact solution of the quenched kinetic spherical model in dimensions $2<d<4$. 

A simple heuristic argument to establish (\ref{3.11}) goes as follows. For widely different times $t\gg s\gg \tau_{\rm mic}$, the asymptotic form of the 
autocorrelator is expressed through the cluster sizes $L$ as $C(t,s)\sim\bigl( L(t)/L(s)\bigr)^{-\lambda}$. If furthermore $t$ is so large that
$L(t)\sim N$ while $s$ is small enough such that still $L(s)\sim s^{1/z}$, the scaling (\ref{3.11}) of the plateau $C_{\infty}^{(2)}$ follows. 

\section{The kinetic spherical model}

Following standard developments \cite{Ronc78,Cugl94,Godr00b,Henk15}, the kinetic spherical model is defined in terms of real spin variables 
$S_{\vec{n}} = S_{\vec{n}}(t)\in\mathbb{R}$ at each lattice site $\vec{n}\in\Lambda\subset\mathbb{Z}^d$, subject to the
{\em spherical constraint} $\sum_{\vec{n}\in\Lambda} S_n^2(t)=|\Lambda|$, where $|\Lambda|=\prod_{j=1}^d N_j$ is the number of sites of the lattice 
$\Lambda\subset\mathbb{Z}^d$. Its dynamics is given by the Langevin equation
\BEQ
\partial_t S_{\vec{n}}(t) = D \Delta_{\vec{n}} S_{\vec{n}}(t) - \mathfrak{z}(t) S_{\vec{n}}(t) + \eta_{\vec{n}}(t)
\EEQ
with the spatial laplacian $\Delta_{\vec{n}}$ and the thermal white noise $\eta_{\vec{n}}=\eta_{\vec{n}}(t)$. It has the first two moments
\BEQ \label{4.2}
\bigl\langle \eta_{\vec{n}}(t) \bigr\rangle = 0 \;\; , \;\;
\bigl\langle \eta_{\vec{n}}(t) \eta_{\vec{m}}(t') \bigr\rangle = 2D T \delta(t-t') \delta_{\vec{n},\vec{m}} 
\EEQ
where $T$ is the bath temperature and $D$ a kinetic coefficient. The Lagrange multiplier $\mathfrak{z}(t)$ is fixed from the
spherical constraint. The Fourier representation
\BEQ \label{4.4}
S_{\vec{n}}(t) = \frac{1}{|\Lambda|} \sum_{k_1=0}^{N_1-1} \cdots \sum_{k_d=0}^{N_d-1} 
\exp\left( 2\pi\II \sum_{j=1}^d \frac{k_j}{N_j} n_j  \right) \wht{S}(t,\vec{k})
\EEQ
achieves the formal solution of the model which reads
\begin{subequations} \label{4.5}
\BEQ \label{4.5a}
\wht{S}(t,\vec{k}) = \wht{S}(0,\vec{k}) \frac{\exp\bigl( -2D \omega(\vec{k})t\bigr)}{\sqrt{g(t)\,}} 
+ \int_0^t \!\D\tau\:  \wht{\eta}(\tau,\vec{k}) \sqrt{\frac{g(\tau)}{g(t)}\,}\, \exp\bigl( -2D\omega(\vec{k})(t-\tau)\bigr)
\EEQ
with the abbreviations (nearest-neighbour interactions assumed) 
\BEQ \label{4.5b}
\omega(\vec{k}) = \sum_{j=1}^d \left( 1 - \cos \frac{2\pi}{N_j} k_j \right) \;\; , \;\;
g(t) = \exp\left( 2 \int_0^t \!\D\tau\: \mathfrak{z}(\tau) \right)
\EEQ
In what follows, we restrict to a totally disordered initial state, such that 
$\bigl\langle S_{\vec{n}}(0)\bigr\rangle =0$ and 
$\bigl\langle S_{\vec{n}}(0) S_{\vec{m}}(0)\bigr\rangle=\delta_{\vec{n},\vec{m}}$. 
In momentum space, the second moments of initial and thermal noises become 
\BEQ \label{4.5c}
\left\langle \wht{S}(0,\vec{k}) \wht{S}(0,\vec{k}')\right\rangle = |\Lambda| \delta_{\vec{k}+\vec{k}',\vec{0}} \;\; , \;\;
\bigl\langle \wht{\eta}(t,\vec{k}) \wht{\eta}(t',\vec{k}')\bigr\rangle 
= 2DT |\Lambda| \delta(t-t') \delta_{\vec{k}+\vec{k}',\vec{0}}
\EEQ
Then the spherical constraint can be cast into a Volterra integral equation for $g=g(t)$
\BEQ \label{4.5d} 
g(t) = f(t) + 2DT \int_0^t \!\D\tau\: g(\tau) f(t-\tau) \;\; , \;\;
f(t) := \frac{1}{|\Lambda|} \sum_{\vec{k}} \exp\left( -4D \omega(\vec{k}) t\right)
\EEQ
\end{subequations}
Here and below, we abbreviate $\sum_{\vec{k}} := \sum_{k_1=0}^{N_1-1} \cdots \sum_{k_d=0}^{N_d-1}$. 
Eqs.~(\ref{4.5}) specify the exact solution of the kinetic spherical model. We are interested in
\begin{enumerate}
\item[{\bf (I)}] the two-time correlation function $\wht{C}(t,s;\vec{k})$ in momentum space, defined by 
\begin{subequations} \label{4.6}
\begin{align}
\left\langle\wht{S}(t,\vec{k})\wht{S}(s,\vec{k}')\right\rangle &=: 
|\Lambda|\delta_{\vec{k}+\vec{k}',\vec{0}}\:\wht{C}(t,s;\vec{k}) \label{4.6a} \\
\wht{C}(t,s;\vec{k}) &= \frac{e^{-2D\omega(\vec{k})(t+s)}}{\sqrt{g(t) g(s)\,}} 
+ 2D T \int_0^{\min(t,s)} \!\!\D\tau\: \frac{g(\tau)}{\sqrt{g(t)g(s)\,}\,}\: e^{-2D\omega(\vec{k})(t+s-2\tau)} \label{4.6b} 
\end{align}
\end{subequations}
and especially the two-time autocorrelator
\BEQ \label{4.7}
C(t,s) := \frac{1}{|\Lambda|} \sum_{\vec{k}} \wht{C}(t,s;\vec{k}) = C(s,t)
\EEQ
\item[{\bf (II)}] the single-time correlator in momentum space $\wht{C}(t;\vec{k}):=\wht{C}(t,t;\vec{k})$, obtained from (\ref{4.6}) by setting $s=t$. 
The time-space correlator reads 
\BEQ \label{4.8} 
C(t;\vec{n}) = \frac{1}{|\Lambda|}\sum_{\vec{k}}\exp\left( 2\pi\II\sum_{j=1}^d \frac{k_j}{N_j} n_j\right) \wht{C}(t;\vec{k})
\EEQ
\end{enumerate}
The well-known bulk critical temperature \cite{Berl52} ($I_0(u)$ is a modified Bessel function \cite{Abra65})
\BEQ \label{5.1}
\frac{1}{T_c(d)} =  \int_0^{\infty} \!\D u\: \left( e^{-2u} I_0(2u) \right)^d
\EEQ
is finite and positive for $d>2$. Explicitly \cite{Cara03,Borw92}
\BEQ
\frac{1}{T_c(3)} = 
\frac{\sqrt{3\,}-1}{192\pi^3}\left( \Gamma\left(\frac{1}{24}\right)\Gamma\left(\frac{11}{24}\right)\right)^2 
\approx 0.25273\ldots
\EEQ

In what follows, we consider a hyper-cubic geometry $\overbrace{N\times\cdots \times N\,}^{d^*\: \mbox{\rm\scriptsize factors}}\times 
\overbrace{\infty \times \cdots \times \infty}^{d-d^*\: \mbox{\rm\scriptsize factors}}$, 
where the first $d^*\leq d$ directions are finite and periodic and the other $d-d^*$ directions are infinite.
We also restrict to $2<d<4$ and rescale the temporal units such that $8\pi D \stackrel{!}{=}1$. 
After a quench from the disordered initial state (\ref{4.5c}) to a temperature $T<T_c(d)$, 
we find in the {\sc fss} limit (\ref{3.3}) (see the appendix for the calculations)

\noindent
{\bf (A)} the single-time temporal-spatial correlator, namely 
\begin{subequations} \label{4.10} 
\begin{align}
C(t;\vec{n}) &= M_{\rm eq}^2 \exp\left( -\pi\sum_{j=1}^d \frac{n_j^2}{t} \right)
\prod_{j=1}^{d^*} 
\frac{\vartheta_3\bigl(\II\pi\frac{N n_j}{t},e^{-\pi Z}\bigr)}{\vartheta_3\bigl(0,e^{-\pi Z}\bigr)}
\label{4.10a} \\
&= M_{\rm eq}^2 \exp\left( -\pi \sum_{j=d^*+1}^d \frac{n_j^2}{t} \right) 
\prod_{j=1}^{d^*} \frac{\vartheta_3\bigl( \pi n_j/N, \exp(-\pi/Z)\bigr)}{\vartheta_3\bigl(0, \exp(-\pi/Z)\bigr)}
\label{4.10b}
\end{align}
\end{subequations}
where $M_{\rm eq}^2 = 1 -T/T_c(d)$ is the squared equilibrium magnetisation and $Z$ was defined in (\ref{3.3}) with $z=2$. Finally, 
$\vartheta_3(z,q)=\sum_{p=-\infty}^{\infty} q^{p^2} \cos(2p z)$ is a Jacobi Theta function 
\cite{Abra65}.\footnote{Analogous expressions of the finite-size scaling functions in terms
of Jacobi Theta functions are known for the 
particle density in several $1D$ reaction-diffusion processes for both periodic and open boundary conditions 
\cite{Krebs94a,Krebs94b} and for the single-time correlator in the 
periodic $1D$ Glauber-Ising model at temperature $T=0$ \cite{Alca94}.} See figure~\ref{fig1B}a for illustration. From (\ref{4.10a}) we identify the
finite-size scaling function $F_C=F_C(\vec{\vro},Z)$ in (\ref{3.2}). The shape of this function is temperature-independent. Indeed, an universal shape
of $F_C$ is expected, since the temperature $T$ should be irrelevant in phase-ordering kinetics \cite{Bray94a}. 

Eq.~(\ref{4.10a}) gives a factorisation of $C(t,\vec{n})=C_{\rm bulk}(t;\vec{n})\cdot C_{\rm red}(t;\vec{n};N)$ into a size-independent `bulk' part and a
`reduced' part which contains the finite-size effects. Because of the identity $\vartheta_3(z+\pi,q)=\vartheta_3(z,q)$, 
it is seen from (\ref{4.10b}) that the correlator repeats periodically when $n_j\mapsto n_j+N$ 
is in the finite directions, as illustrated in the inset of figure~\ref{fig1B}a. 
For $Z$ large enough\footnote{Actually for $Z\gtrsim 25$, which in physical units corresponds to $L(t)\lesssim 5 N$.} 
the central peak of the correlator around
$\vec{n}=\vec{0}$ decays as in the bulk with a length scale $L(t)\sim t^{1/2}$ such that the system decomposes into separate and independent 
clusters of linear size $L(t)$, as expected. 
The bulk gaussian decay $\sim e^{-\vec{n}^2/t}$, rather than an exponential $\sim e^{-|\vec{n}|/\sqrt{t\,}}$, 
is a peculiar property of the spherical model which distinguishes it from the Ising universality class.

\noindent
{\bf (B)} the two-time autocorrelator, for all $T<T_c(d)$, reads 
\begin{subequations} \label{4.11}
\begin{align} \label{4.11a} 
C(ys,s) &= M_{\rm eq}^2 \left( \frac{2 \sqrt{y\,}}{1+y}\right)^{d/2} \left( 
\frac{\vartheta_3\bigl(0,\exp(-\pi\frac{2Z}{1+1/y})\bigr)^2}{\vartheta_3\bigl(0,\exp(-\pi{Z})\bigr)
\vartheta_3\bigl(0,\exp(-\pi Zy)\bigr)}\right)^{d^*/2} \\
&= M_{\rm eq}^2  \left( \frac{2 \sqrt{y\,}}{1+y}\right)^{(d-d^*)/2} \left(
 \frac{\vartheta_3\bigl(0,\exp(-\pi\frac{1+1/y}{2Z})\bigr)^2}{\vartheta_3\bigl(0,\exp(-\pi/Z)\bigr)
\vartheta_3\bigl(0,\exp(-\pi /Zy)\bigr)}\right)^{d^*/2}  \label{4.11b} 
\end{align}
\end{subequations}
as illustrated in figure~\ref{fig1B}b. We identify from (\ref{4.11a}) the finite-size scaling function $f_C=f_C(y,Z)$ in (\ref{3.2}), 
whose shape is once more temperature-independent. 
As above for the single-time correlator, (\ref{4.11a}) displays a natural factorisation into the bulk two-time autocorrelator 
$C_{\rm bulk}(y s,s)=M_{\rm eq}^2 \left( \frac{2\sqrt{y\,}}{1+y}\right)^{d/2}$ and 
a `reduced' factor which alone contains all finite-size effects. 
Eq.~(\ref{4.11a}) shows that for $Z\gg 1$, finite-size corrections with respect to the
bulk behaviour are exponentially small. On the other hand, eq.~(\ref{4.11b}) shows that for $Z\ll 1$, 
the system behaves effectively as if it had only $d-d^*$ dimensions, 
up to exponentially small corrections.\footnote{Finite-temperature and finite-time effects merely give a corrective factor  $1+{\rm O}(T s^{1-d/2})$,
negligible for large waiting times $s\to\infty$, if $d>2$.}  

Having verified the generic finite-size scaling forms (\ref{3.2}), we now test the validity of the finite-size scaling predictions (\ref{3.12}) 
for the plateau values $C^{(2)}_{\infty}$. To be specific, we consider a fully finite system, with $d^*=d$. 
Fix the system size $N$ and the waiting time $s$ and consider the
changes in $y=t/s$ by varying the observation time $t$. 
Physically, finite-size effects will be felt first by the larger length $L(t)\sim t^{1/2}$. Since $t\gg s$, we expect that $L(t)\gg L(s)$. 
The limit $y\gg 1$ is realised by taking $t\gg 1$. With the identity $\vartheta_3\bigl(0,e^{-\pi y}\bigr) = y^{-1/2}\, \vartheta_3\bigl(0,e^{-\pi/y}\bigr)$,
we have 
\BEQ \label{4.12theta}
C(t,s) = M_{\rm eq}^2 \left( \frac{t \bigl(s/t\bigr)^{1/2}}{(t+s)/2}\right)^{d/2} 
\left( \frac{ \bigl( 2 \frac{N^2}{t+s}\bigr)^{-1/2} \vartheta_3\bigl(0, \exp(-\frac{\pi}{2}\frac{t+s}{N^2})\bigr)}{\sqrt{
\bigl( \frac{N^2}{t}\bigr)^{-1/2}\vartheta_3\bigl(0, \exp(-\pi\frac{t}{N^2})\bigr)
\vartheta_3\bigl(0, \exp(-\pi\frac{N^2}{s})\bigr)\,}\,}\right)^{d^*}
\EEQ
For $N^2/s$ finite but large enough (such that the plateaux in figure~\ref{fig1B}b are reached) , the last of the Theta functions in 
(\ref{4.12theta}) is very close to unity. Because of the condition $t/N^2\gg 1$, the other two Theta-functions in (\ref{4.12theta}) are also close to unity. 
Up to constants, we obtain
\BEQ
C(t,s) \stackrel{t\to\infty}{\sim} \left(\frac{s}{t}\right)^{d/4} 
       \left( \left( \frac{t+s}{N^2}\right)^{1/2} \left( \frac{N^2}{t}\right)^{1/4} \mbox{\rm cste.} \right)^{d^*} 
\sim \left(\frac{s}{t}\right)^{d/4} \left( \frac{t(1+s/t)}{t^{1/2}}\right)^{d^*/2} \left( N^{-2\frac{1}{2} + 2\frac{1}{4}}\right)^{d^*} 
\EEQ
Finally, now admitting a fully finite system such that $d=d^*$, we have (for $2<d<4$) 
\BEQ
C(t,s) \sim \left(\frac{s}{t}\right)^{d/4} t^{d/4} N^{-d/2} = s^{d/4} N^{-d/2}
\EEQ
which in view of the well-known results $\lambda=d/2$ \cite{Godr00b} and $z=2$ \cite{Bray94a} 
does indeed reproduce of (\ref{3.11}), or (\ref{3.12}) if either $s$ or $N$ is kept fixed.

\noindent 
{\bf (C)} Characteristic time-dependent length scales $L(t)$ of the ordered clusters can be measured as second moments of the single-time correlator
\BEQ \label{5.8}
L^2(t) := \frac{\sum_{\vec{n}} \vec{n}^2 C(t;\vec{n})}{\sum_{\vec{n}} C(t;\vec{n})}
\EEQ
Precise expressions follow from (\ref{4.10a}) once the range of summation of the distances $|\vec{n}|$ 
is fixed. For example, if one measures the distances along one of the coordinate axes of one of the
infinite directions, one obtains the {\em `transverse'} length scale $L_{\perp}^2(t) = 4D t$, as for a fully infinite system \cite{Ebbi08}. 
On the other hand, if the distances are measured along the coordinates axes of one of the finite directions, we find a 
{\em `longitudinal'} length scale, which reads for sufficiently thick films, and in agreement with (\ref{3.4})
\BEQ \label{5.9}
L_{\|}^2(t) = \frac{1}{\pi} t f_{L}(Z) \;\; , \;\;
f_{L}(Z) = \frac{\pi}{6} Z \left( 1 + \frac{12}{\pi^2}\sum_{p=1}^{\infty} \frac{(-1)^p}{p^2}\, e^{-\pi p^2/Z}\right)
\simeq \left\{ \begin{array}{ll} \frac{\pi}{6} Z & \mbox{\rm ;~ if $Z\ll 1$} \\[0.12truecm]
                                           1     & \mbox{\rm ;~ if $Z\gg 1$}  
   \end{array} \right.  
\EEQ
The scaling function $f_L$ is temperature-independent. 
This describes the cross-over shown in figure~\ref{fig2B}, such that for $Z=N^2/t$ small enough, 
we obtain saturation at $L_{\|}^2(t)\to L_{\infty}^2 \sim N^2$, 
but on the other hand one has $L_{\|}^2(t)\sim t$ of an effectively infinite system for $Z$ large enough.

\section{{\it Ad conclusio}}

We studied finite-size scaling in the ageing relaxation of 
phase-ordering kinetics after a quench from a disordered initial state into
the two-phase coexistence regime with temperature $0<T<T_c$. 
The finite-size scaling {\em ansatz} (\ref{3.1}) is the natural extension of dynamic finite-size scaling at equilibrium \cite{Suzuki77}. 
Phenomenologically, the observations to be gleaned from figure~\ref{fig1B} for the
single-time and two-correlations and figure~\ref{fig2B} for the characteristic length scale are captured by the finite-size scaling forms
(\ref{3.2}). The {\em form} of the associated scaling functions is temperature-independent, which confirms the expectation that the temperature should be
irrelevant in phase-ordering kinetics \cite{Bray94a}. From these, the finite-size scaling (\ref{3.4}) 
for the length scale $L_{\|}(t)$ and especially (\ref{3.12}) for the plateaux $C^{(2)}_{\infty}$ in the
two-time autocorrelator of a fully finite system were derived. We checked that these predictions are fully bourne out in the phase-ordering of the
exactly solved kinetic spherical model, for $2<d<4$ dimensions. 

Clearly, several open questions remain, including:
\begin{enumerate}
\item Do the {\sc fss} predictions (\ref{3.2},\ref{3.4},\ref{3.12}) also hold for other universality classes~? For kinetic Ising models with
either short-ranged or long-ranged interactions, detailed tests on all these have been carried out recently and will be reported elsewhere \cite{Chris22}. 
\item Although the discussion was entirely formulated here in terms of classical dynamics, 
a finite-size scaling ansatz such as (\ref{3.2}) should {\it a priori} 
also work for relaxations in quantum systems, either closed or open.  
\item Our analysis is restricted to below the upper critical dimension $d<d_c$. At equilibrium, it is well-known that dangerous irrelevant variables
lead to essential modifications of the finite-size scaling ansatz (\ref{3.1},\ref{3.2}) 
\cite{Bind85,Priv90,Kenn14,Flor15,Flor16,Grim17,Berc22}. Such modifications should also become necessary for the dynamics. 

Considerations of this kind might become crucial either for long-range interactions, where $d_c$ is lowered with respect to the value 
$d_c^{({\rm short})}=4$ of short-ranged systems or
else for $d$-dimensional {\em quantum} systems (possibly with long-ranged interactions as well), 
for which at least the equilibrium quantum phase transitions at $T=0$ are known to be in the same universality class as the corresponding
$(d+\theta)$-dimensional classical universality class at finite temperature, where the anisotropy exponent $\theta\geq 1$ \cite{Sach11}. 
\item From figure~\ref{fig1B}b it appears that finite-size effects might create a spurious regime where the autocorrelator 
$C(ys,s)\sim y^{-\lambda_{\rm eff}}$ might look
algebraic in a certain window; but rather the system already is the transition region 
between the rapid fall-off after having left the infinite-size behaviour of $C_{\rm bulk}(ys,s)$ and the turn-around
towards the saturation plateau $C^{(2)}_{\infty}$. Since $\lambda_{\rm eff}>\lambda$, 
not recognising this effect carries the risk of systematic over-estimation of the auto-correlation exponent $\lambda$, in simulations or in experiments. 
\item One may generalise dynamical {\sc fss} to critical quenches and to two-time response functions as well. The theory and numerical tests thereof 
will be presented elsewhere \cite{Chris22}. 
\item Can one use (\ref{3.12}) to devise improved methods for the measurement of $\lambda$~? 
\end{enumerate}


\appsektion{Analytical derivations}
The exact solution of the kinetic spherical model at $T<T_c(d)$, starting from (\ref{4.5}), is described. 

\subsection{Spherical constraint}

The Volterra integral equation (\ref{4.5d}) gives the long-time behaviour of $g(t)$ in a large, but finite system, as follows.  
The first part retraces the steps used at equilibrium \cite{Singh85,Singh87}, with the notation adapted for dynamics. 
The second part gives the new ingredients needed for non-equilibrium dynamics. \\

\noindent
{\bf 1.} Through a Laplace transform we formally solve (\ref{4.5d}) 
\BEQ \label{A.1}
\lap{g}(p) = \mathscr{L}(g)(p) := \int_0^{\infty} \!\D t\: e^{-pt} g(t) = \frac{\lap{f}(p)}{1-2D T \lap{f}(p)}
\EEQ
Standard Tauberian theorems \cite{Fell71} relate the behaviour of $\lap{g}(p)$ in the $p\to 0$ 
limit to the asymptotic long-time behaviour of $g(t)$ for $t\to\infty$. 
One needs the leading terms of $\lap{f}(p)$ as $p\to 0$. Recall the generalised Poisson re-summation formula \cite{Wilt30}
\BEQ \label{A.2}
\sum_{n=a}^{b} f(n) = \sum_{q=-\infty}^{\infty} \int_a^b \!\D x\: e^{2\pi\II q x} f(x) + \demi f(a) + \demi f(b)
\EEQ
and use this to deduce the important identity, for $m\in\mathbb{Z}$ and $x\in\mathbb{R}$
\BEQ \label{A.3} 
\sum_{k=0}^{N-1} \exp\left(  \frac{2\pi\II}{N} k m + x \cos\frac{2\pi k}{N} \right) 
= N \sum_{q=-\infty}^{\infty} I_{qN+m}(x)
\EEQ
where $I_n(x)$ is a modified Bessel function \cite{Abra65}. 

Now, one writes as in \cite{Singh85}, using eq.~(\ref{A.3}) with $m=0$ in the second line $d$ times
\BEA
2D \lap{f}(p) &=&  \frac{2D}{|\Lambda|} \sum_{\vec{k}} \int_0^{\infty} \!\D t\: 
\exp\left[ -\left( p +4D \sum_{j=1}^d \left( 1 - \cos\frac{2\pi}{N_j} k_j \right) \right) t \right] 
\nonumber \\
&=& 2D \int_0^{\infty} \!\D t\: e^{-(p+4Dd)t} \sum_{q_1,\ldots , q_d\in\mathbb{Z}} \prod_{j=1}^d I_{N_j q_j}(4D t)
\nonumber \\
&=& \demi \int_0^{\infty} \!\D u\: e^{-\demi \phi u} \left( e^{-u} I_0(u) \right)^d 
 + \demi {\sum_{\vec{q}\in\mathbb{Z}^d}}' \int_0^{\infty} \!\D u\: e^{-\demi\phi u} 
\prod_{j=1}^{d} \left( e^{-u} I_{N_j q_j}(u) \right)
\EEA
where one sets $\phi := p/2D$. In the last line, the bulk contribution which arises from $\vec{q}=\vec{0}$, is separated
from the finite-size terms which have $\vec{q}\ne\vec{0}$ (indicated by ${\sum}'$). 

In what follows, restrict throughout to dimensions $2<d<4$. First, standard techniques \cite{Barb73,Brez82,Luck85,Godr00b} 
give the leading order of the Watson function $W_d(\phi)$ for $\phi\ll 1$, as follows
\BEA
W_d(\phi) &:=& \demi \int_0^{\infty} \!\D u\: e^{-\demi \phi u} \left( e^{-u} I_0(u) \right)^d \nonumber \\
&\simeq& W_d(0) - (4\pi)^{-d/2} \left| \Gamma\left(1-\frac{d}{2}\right)\right| \phi^{(d-2)/2} \left( 1 + {\rm o}(\phi)\right)
\EEA
with an implied analytic continuation in $d$. 
Next, the finite-size terms are evaluated in the hyper-cubic geometry, such that the 
first $d^*$ dimensions are finite ($0< d^*\leq d$), with periodic boundary conditions (for simplicity, set $N_j=N$ for all $j=1,\ldots,d^*$). 
The remaining $d-d^*$ dimensions are assumed to be infinite, formally $N_j=\infty$. 
With the asymptotic identity \cite{Singh85}
$I_{\nu}(x)=(2\pi x)^{-1/2} e^{x-\nu^2/2x}\bigl( 1 + {\rm O}(1/x)\bigr)$ one has 
\BEA
\lefteqn{ \hspace{-0.2cm}
\demi  \int_0^{\infty} \!\D u\: e^{-\demi\phi u} \prod_{j=1}^{d} \left( e^{-u} I_{N_j q_j}(u) \right) \;\simeq\; 
\demi \int_0^{\infty} \!\D u\: e^{-\demi\phi u} \bigl( 2\pi u\bigr)^{-d/2} 
\prod_{j=1}^{d^*} e^{ -(N q_j)^2/2u} ~~~~
} \nonumber \\
&=& (4\pi)^{-d/2} \phi^{d/2-1} \int_0^{\infty} \!\D v\: v^{-d/2} 
\exp\left( -v - \frac{1}{v}\frac{\phi}{4} \sum_{j=1}^{d^*} N^2 q_j^2 \right) 
\nonumber \\
&=& \frac{2}{(4\pi)^{d/2}} \left( \frac{2\psi}{N}\right)^{d-2} 
\left( \frac{1}{\psi |\vec{q}|}\right)^{(d-2)/2} K_{(d-2)/2} \bigl( 2\psi |\vec{q}|\bigr)
\label{A.6}
\EEA
with the {\em thermo-geometric parameter} $\psi := \demi N \phi^{1/2}$, the short-hand 
$|\vec{q}|^2 := \sum_{j=1}^{d^*} q_j^2$,
the other modified Bessel function $K_{\nu}(x)$ \cite{Abra65} and where the identity \cite{Singh85} 
\BEQ
\int_0^{\infty} \!\D x\: x^{\nu-1} e^{-\beta x - \alpha/x} = 2 \left(\frac{\alpha}{\beta}\right)^{\nu/2}
K_{\nu}\bigl( 2\sqrt{\alpha\beta\,}\,\bigr)
\EEQ
was used in the last line. In the infinite directions, only the terms with $q_j=0$ contribute in (\ref{A.6}), 
for $j=d^*+1,\ldots,d$. The final result of the first part is, for $2<d<4$ \cite{Singh85,Singh87}
\BEQ \label{A.8}
2D \lap{f}(p) = W_d(0) - \frac{1}{(4\pi)^{d/2}} \left( 
\left| \Gamma\left(1-\frac{d}{2}\right)\right| - 2 {\sum_{\vec{q}\in\mathbb{Z}^{d^*}}}'
\frac{K_{(d-2)/2}(2\psi|\vec{q}|)}{(\psi |\vec{q}|)^{(d-2)/2}}\right) \left(\frac{2\psi}{N}\right)^{d-2} + \ldots
\EEQ

\noindent
{\bf 2.} We define the abbreviation
\BEQ
H_{\alpha}(\psi) := \frac{1}{(4\pi)^{d/2}} \left( \left| \Gamma\left(-\alpha\right)\right| - 2 {\sum_{\vec{q}^*}}'
\frac{K_{\alpha}(2\psi|\vec{q}|)}{(\psi |\vec{q}|)^{\alpha}}\right)
\EEQ
where $\sum_{\vec{q}^*} = \sum_{\vec{q}\in\mathbb{Z}^{d^*}}$ is only extended over the finite directions. 
In the spherical model, the equilibrium magnetisation $M_{\rm eq}^2 = 1 -T/T_c$, 
where the critical temperature $1/T_c=W_d(0)$ \cite{Berl52,Barb73,Singh85,Ronc78,Godr00b}. 
For quenches to $T<T_c$ one has $M_{\rm eq}^2>0$. 
Then, using (\ref{A.1}) and (\ref{A.8}) 
\BEA
\lap{g}(p) &\simeq& \frac{1}{2D} 
\frac{W_d(0) - H_{(d-2)/2}(\psi) 
\bigl(\frac{2\psi}{N}\bigr)^{d-2} + \ldots}{1-TW_d(0) +T H_{(d-2)/2}(\psi) \bigl(\frac{2\psi}{N}\bigr)^{d-2} + \ldots} 
\nonumber \\
&\simeq& \frac{1}{2D T_c}\frac{1}{M_{\rm eq}^2} 
         - \frac{1}{2D}\frac{1}{M_{\rm eq}^4} H_{(d-2)/2}(\psi) \left(\frac{2\psi}{N}\right)^{d-2} + \ldots
\nonumber \\
&=& \frac{1}{2D T_c}\frac{1}{M_{\rm eq}^2} - \frac{1}{2D M_{\rm eq}^4}\frac{|\Gamma(1-d/2)|}{(4\pi)^{d/2}}\left(\frac{p}{2D}\right)^{(d-2)/2} 
\nonumber \\
& & +\frac{2}{2D M_{\rm eq}^4}\frac{1}{(4\pi)^{d/2}} \left(\frac{p}{2D}\right)^{(d-2)/4} {\sum_{\vec{q}^*}}' 
\left( \frac{N|\vec{q}|}{2}\right)^{(2-d)/2} K_{(d-2)/2}\left(\frac{N|\vec{q}|}{\sqrt{2D\,}} p^{1/2}\right) ~~
\label{A.10}
\EEA
gives the leading terms of $\lap{g}(p)$ for small values of $p$. The first two of these terms are the bulk contributions,
while the remaining ones give the leading finite-size effects. 

The leading long-time behaviour of $g(t)$ is then obtained via the identities \cite{Prud5} 
\begin{subequations}
\begin{align}
\mathscr{L}^{-1}\bigl( p^{\nu/2} K_{\nu}(2ap^{1/2}) \bigr)(t) &= \demi \frac{a^{\nu}}{t^{\nu+1}}\, e^{-a^2/t} 
 \\
\mathscr{L}^{-1}\bigl( p^{-\nu}\bigr)(t) &= \frac{1}{\Gamma(\nu)} t^{\nu-1}
\end{align}
\end{subequations}
and we find, where from now on both $d$ and $d^*$ can be considered as continuous parameters 
\BEA
g(t) &=& \frac{1}{2D  T_c}\frac{1}{M_{\rm eq}^2} \delta(t) + \frac{1}{M_{\rm eq}^4}\frac{1}{(8\pi D t)^{d/2}} 
+ \frac{1}{M_{\rm eq}^4(8\pi D t)^{d/2} } {\sum_{\vec{q}^*}}' e^{-\pi \frac{N^2}{8\pi D t} |\vec{q}|^2} 
\nonumber \\
&=& \frac{1}{2D  T_c}\frac{1}{M_{\rm eq}^2}\, \delta(t) 
+ \frac{(8\pi D t)^{-d/2}}{M_{\rm eq}^4}\, \vartheta_3\left(0,\exp\left(-\pi \frac{N^2}{8\pi D t}\right)\right)^{d^*} 
\label{A.12}
\EEA
with the Jacobi Theta function $\vartheta_3$ \cite{Abra65}, which obeys the functional identity 
\BEQ \label{A.13}
\vartheta_3\bigl(0,e^{-\pi y}\bigr) = y^{-1/2}\, \vartheta_3\bigl(0,e^{-\pi/y}\bigr)
\EEQ
\begin{figure}[tb]
\begin{center}
\includegraphics[width=.48\hsize]{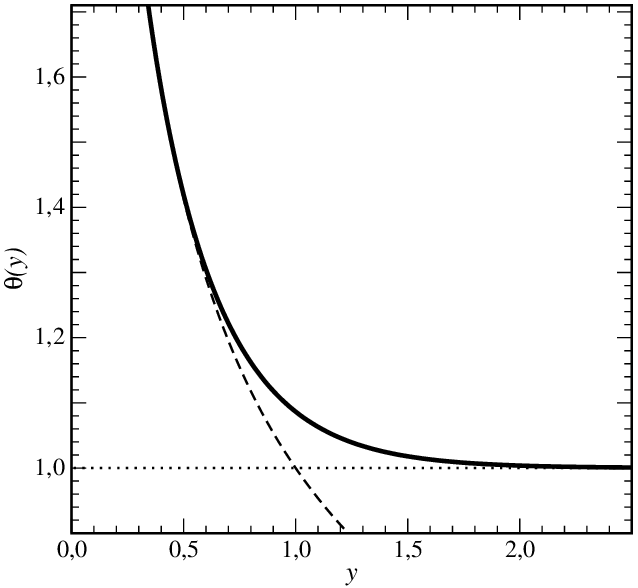} 
\end{center}
\caption[fig4]{\small The function $\theta(y):=\vartheta_3\bigl(0,e^{-\pi y}\bigr)$ and its cross-over between the regimes
where $y\gg 1$ and $\theta(y)\simeq 1$ (dotted line) and $y\ll 1$ and $\theta(y)\simeq y^{-1/2}$ (dashed line). 
\label{fig4} }
\end{figure}
Figure~\ref{fig4} illustrates the rapid cross-over (essentially in the interval $\demi\lesssim y\lesssim 2$) between the two
asymptotic regimes. Therefore,  
we have the following asymptotic limits, for $2<d<4$ and $T<T_c$ 
\BEQ \label{A.14}
g(t) =  \frac{1}{2D  T_c}\frac{1}{M_{\rm eq}^2} \delta(t) + 
\left\{ \begin{array}{lll} 
 \frac{(8\pi D t)^{-d/2}}{M_{\rm eq}^4} & \mbox{\rm ~;~~ if $N^2/t\gg 1$} & \mbox{\rm ~~ infinite-size system} \\[0.12truecm]
 \frac{(8\pi D t)^{-(d-d^*)/2}}{M_{\rm eq}^4} N^{-d^*} & \mbox{\rm ~;~~ if $N^2/t\ll 1$} & \mbox{\rm ~~ finite-size system}
\end{array} \right.
\EEQ
This shows that the long-time behaviour of the spherical constraint in a finite geometry is effectively ($d-d^*$)-dimensional. 
The singular terms in (\ref{A.12},\ref{A.14}) will become very important for the calculation of the correlators, as we shall see below. 

Eq.~(\ref{A.12}) is the main result of this sub-section.

\subsection{Two-time autocorrelator} 

We decompose in (\ref{A.12}) $g(t)=g_{\rm sing}(t) + g_{\rm reg}(t)$, where $g_{\rm sing}(t) = \frac{1}{2DT_c}\frac{1}{M_{\rm eq}^2} \delta(t)$. 
In momentum space, with the convention $t>s$, we have from (\ref{4.6}), for large times, the decomposition
\BEA
\wht{C}(t,s;\vec{k}) &=& \frac{e^{-2D\omega(\vec{k})(t+s)}}{\sqrt{g_{\rm reg}(t) g_{\rm reg}(s)\,}} 
\left\{ 1 + \frac{2D T}{2D T_c} \frac{1}{M_{\rm eq}^2} \int_0^s \!\D\tau\: \delta(\tau) e^{2D\omega(\vec{k})2\tau} 
+ 2D T \int_0^s \!\D\tau\: g_{\rm reg}(\tau) e^{2D\omega(\vec{k})2\tau} \right\}
\nonumber \\
&=& \frac{1}{M_{\rm eq}^2}\frac{e^{-2D\omega(\vec{k})(t+s)}}{\sqrt{g_{\rm reg}(t) g_{\rm reg}(s)\,}} 
+2DT \int_0^s \!\D\tau\: \frac{g_{\rm reg}(\tau)}{\sqrt{g_{\rm reg}(t) g_{\rm reg}(s)\,}}\, e^{-2D\omega(\vec{k})(t+s-2\tau)}
\label{A.16}
\EEA
for all temperatures $T<T_c$. With (\ref{4.7}), this gives the two-time autocorrelator 
$C(t,s) = C^{[1]}(t,s) + C^{[2]}(t,s)$. 
The first term in (\ref{A.16}) leads to
\BEA
C^{[1]}(t,s) &=& \frac{ |\Lambda|^{-1} M_{\rm eq}^{-2}}{\sqrt{g_{\rm reg}(t) g_{\rm reg}(s)\,}} 
\sum_{\vec{k}} \exp\left[ - 2D\sum_{j=1}^d \left( 1 - \cos \frac{2\pi}{N_j} k_j\right) (t+s) \right] 
\nonumber \\
&=& \frac{M_{\rm eq}^{-2}}{\sqrt{g_{\rm reg}(t) g_{\rm reg}(s)\,}} \prod_{j=1}^d \sum_{q_j\in\mathbb{Z}} 
e^{-2D(t+s)} I_{N_j q_j}(2D(t+s)) 
\nonumber \\
&\simeq& \frac{M_{\rm eq}^{-2}}{\sqrt{g_{\rm reg}(t) g_{\rm reg}(s)\,}} \frac{1}{(4\pi D(t+s))^{d/2}}\prod_{j=1}^{d^*} \sum_{q_j\in\mathbb{Z}} 
\exp\left[-\frac{(N q_j)^2}{4D(t+s)}\right] \biggl( 1 + {\rm O}\bigl((t+s)^{-1}\bigr) \biggr)
\nonumber \\
&=& M_{\rm eq}^2 \left( \frac{t^{d/2} s^{d/2}}{\bigl( (t+s)/2\bigr)^{d}}\right)^{1/2} 
\left( \frac{\vartheta_3\bigl(0,\exp(-\pi \frac{N^2}{4\pi D(t+s)})\bigr)}{\sqrt{\vartheta_3\bigl(0,\exp(-\pi \frac{N^2}{8\pi D t})\bigr)\:
\vartheta_3\bigl(0,\exp(-\pi \frac{N^2}{8\pi D s})\bigr)}\,}\right)^{d^*} ~~~~
\label{4.17}
\EEA
where in the first line (\ref{A.3}) with $m=0$ was used once more. 
In the second line, we use the asymptotic expansion of the modified Bessel function $I_n(x)$. 
In the third and forth lines, $g_{\rm reg}(t)$ was 
inserted with $N_j=N$ for $j=1,\ldots,d^*$ from (\ref{A.12}) 
and the sums in the same line were expressed in terms of the Jacobi Theta function $\vartheta_3$.
Both $d$ and $d^*$ can be taken as continuous variables. 

The second term in (\ref{A.16}) can be expressed as a convolution 
\BEQ
C^{[2]}(ys,s) = \frac{2D T}{\sqrt{g_{\rm reg}(ys) g_{\rm reg}(s)\,}}\,
\mathscr{L}^{-1}\left( \lap{g_{\rm reg}}(p) 
\left(\, \overline{\bigl[ e^{-4D( (y+1)s/2 )} I_0(4D(y+1)s/2) \bigr]^d}\, \right)(p) \right)(s) 
\EEQ
For $s\to\infty$, a Tauberian theorem relates the leading behaviour to the one of the Laplace transform at 
$p\to 0$ \cite{Fell71}. 
In turn, the behaviour of the two factors should be dominated by the long-time behaviour of the original functions. 
Therefore, one expects the leading contribution to be of the order ($g_0$ is the amplitude of $g_{\rm reg}(\tau)$) 
\BEA
C^{[2]}(ys,s) &\simeq&  \frac{2D T}{\sqrt{g_{\rm reg}(ys) g_{\rm reg}(s)\,}}\, 
\int_0^s \!\D\tau\: g_0 \tau^{-d/2} \left( 8\pi D \frac{y+1}{2} (t+s-2\tau) \right)^{-d/2} 
\nonumber \\
&\simeq& 2D T (y s^2)^{d/4} s^{1-d} \int_0^1 \!\D v\: v^{-d/2}\, \biggl( 4\pi D (y+1) (y+1-2v) \biggr)^{-d/2} 
\nonumber \\
&=& {\rm O}\bigl(T s^{1-d/2}\bigr)
\EEA
up to an $s$-independent amplitude. 
For $d>2$, $C^{[2]}(ys,s)$ is negligible in the scaling limit where $s\to \infty$. Hence
for all temperatures $T<T_c$, the leading term of the autocorrelator is $C(t,s) = C^{[1]}(t,s)$. 

Finally, introducing the scaling variables $Z$ and $y$ in (\ref{A.16}), and with the scaling $8\pi D\stackrel{!}{=}1$, 
we arrive at (\ref{4.11a}). With (\ref{A.13}), the equivalent form (\ref{4.11b}) is obtained. 

\subsection{Single-time correlator} 

We re-use the decomposition $g(t)=g_{\rm sing}(t)+g_{\rm reg}(t)$ from above. In momentum space, we decompose 
$\wht{C}(t;\vec{k}) = \wht{C}^{[1]}(t;\vec{k}) + \wht{C}^{[2]}(t;\vec{k})$ and have for all $T<T_c$
\BEA
\wht{C}(t;\vec{k}) &=& \frac{e^{-4D\omega(\vec{k})t}}{g_{\rm reg}(t)} +
\frac{2DT}{g_{\rm reg}(t)} \int_0^t \!\D\tau\: \left( \frac{1}{2DT_c}\frac{1}{M_{\rm eq}^2}\delta(\tau) + g_{\rm reg}(\tau)\right)
e^{-4D\omega(\vec{k})(t-\tau)} 
\nonumber \\
&=& \frac{e^{-4D\omega(\vec{k})t}}{M_{\rm eq}^2\, g_{\rm reg}(t)} + 2DT \int_0^t \!\D\tau\:
\frac{g_{\rm reg}(\tau)}{g_{\rm reg}(t)}\, e^{-4D\omega(\vec{k})(t-\tau)}
\EEA
Herein, the first term is analysed as follows
\BEA
C^{[1]}(t;{\vec{n}}) &=& \frac{|\Lambda|^{-1}}{M_{\rm eq}^2\, g_{\rm reg}(t)} \sum_{\vec{k}} 
\exp\left[ \sum_{j=1}^d \frac{2\pi \II}{N_j} k_j n_j - 4D\left( 1 -\cos\frac{2\pi}{N_j} k_j \right) t \right] 
\nonumber \\
&=& \frac{e^{-4D d t}}{M_{\rm eq}^2\, g_{\rm reg}(t)} \sum_{\vec{q}\in\mathbb{Z}^d} \prod_{j=1}^d I_{N_j q_j + n_j}(4Dt) 
\nonumber \\
&\simeq& \frac{M_{\rm eq}^2}{\vartheta_3\bigl(0,e^{-\pi N^2/(8\pi Dt)}\bigr)^{d^*}} 
\prod_{j=1}^d \sum_{q_j\in\mathbb{Z}} e^{-(q_j N_j +n_j)^2/(8Dt)} 
\label{A.21}
\EEA
where first the full identity (\ref{A.3}) is used $d$ times, then the asymptotic form of the modified Bessel function
$I_n(x)$ is used for $x\gg 1$ and finally, in the chosen finite-size geometry, the asymptotic form (\ref{A.12}) is inserted.
The product over the sums in the last line of  (\ref{A.21}) is evaluated as follows: (i) in the $d-d^*$ infinite directions
where formally $N_j=\infty$, only the terms with $q_j=0$ contribute and lead to a factor
$\exp\bigl[ - \frac{1}{8Dt}\sum_{j=d^*+1}^d n_j^2\bigr]$. (ii) the $d^*$ finite directions with $N_j=N$ produce
$d^*$ factors, each of the form
\BEQ
\sum_{q_j\in\mathbb{Z}} \exp\left[ -\frac{(q_j N + n_j)^2}{8Dt}\right] = e^{-n_j^2/(8Dt)} 
\sum_{q_j\in\mathbb{Z}} \exp\left[ - \frac{N n_j}{4Dt} q_j - \frac{N^2}{8Dt} q_j^2 \right]
\EEQ
With the identity 
\BEA \label{A.22}
e^{-n_j^2/(8Dt)}\,  \vartheta_3\left(\II\pi \frac{N n_j}{8\pi Dt}, e^{-\pi N^2/(8\pi D t)} \right)
&=&  \frac{\sqrt{8\pi Dt\,}}{N} \vartheta_3\left( \pi \frac{n_j}{N}, e^{-\pi(N^2/(8\pi Dt))^{-1}} \right)
\EEA
we finally obtain (and used again (\ref{A.13})) 
\BEA 
C^{[1]}(t;{\vec{n}}) &=& M_{\rm eq}^2 \exp\left( -\pi\sum_{j=1}^d \frac{n_j^2}{8\pi Dt} \right)
\prod_{j=1}^{d^*} 
\frac{\vartheta_3\bigl(\II\pi\frac{N n_j}{8\pi Dt},e^{-\pi N^2/(8\pi Dt)}\bigr)}{\vartheta_3\bigl(0,e^{-\pi N^2/(8\pi Dt)}\bigr)}
\nonumber \\
&=& M_{\rm eq}^2 \exp\left( -\pi\sum_{j=d^*+1}^d \frac{n_j^2}{8\pi Dt} \right)
\prod_{j=1}^{d^*} 
\frac{\vartheta_3\bigl(\pi\frac{n_j}{N},e^{-\pi (N^2/(8\pi Dt))^{-1}}\bigr)}{\vartheta_3\bigl(0,e^{-\pi (N^2/(8\pi Dt))^{-1}}\bigr)}
\label{A.23}
\EEA
The second term can be re-written as follows 
\BEQ
C^{[2]}(t;{\vec{n}}) = 2DT \sum_{\vec{q}} \int_0^t \!\D\tau\: \frac{g_{\rm reg}(\tau)}{g_{\rm reg}(t)} 
\prod_{j=1}^d e^{-4D(t-\tau)} I_{q_jN_j + n_j}(4D(t-\tau))
\EEQ
and takes the form of a convolution. For large times $t\to\infty$, we estimate this asymptotically by appealing to Tauberian theorems \cite{Fell71}. 
Then the leading term should become
\BEA
C^{[2]}(t;{\vec{n}}) &\simeq& \frac{2DT}{(8\pi D)^{d/2}}
\int_0^t\!\D\tau\: t^{-d/2} \left(1-\frac{\tau}{t}\right)^{-d/2} \exp\left[-\pi \sum_{j=1}^d \frac{n_j^2}{8\pi D(t-\tau)}\right]  \nonumber \\
& &\times \prod_{j=1}^{d^*} \frac{ \vartheta_3\bigl( \II\pi \frac{N n_j}{8\pi D (t-\tau)}, e^{-\pi N^2/(8\pi D(t-\tau))}\bigr)\, 
\vartheta_3\bigl( 0,  e^{-\pi N^2/(8\pi D\tau)}\bigr)}{  
\vartheta_3\bigl( 0,  e^{-\pi N^2/(8\pi D t)}\bigr)}  \nonumber \\
&\sim& {\rm O}(T t^{1-d/2})
\label{A.25} 
\EEA
which becomes negligible in the long-time limit $t\to \infty$ for $d>2$. 

Therefore, in the long-time limit $t\to\infty$, $C(t;{\vec{n}})=C^{[1]}(t;{\vec{n}})$. Introducing the scaling variables (\ref{3.4}) 
into (\ref{A.23}), and re-using (\ref{A.13},\ref{A.22}) and scaling $8\pi D\stackrel{!}{=}1$, we arrive at eqs.~(\ref{4.10}). 

\subsection{Characteristic length} 

The characteristic lengths $L(t)$ are defined from (\ref{5.8}), with the single-time correlator given by (\ref{A.23}). 
If the distances are calculated along the coordinates axes in one of the $d^*$ finite directions, i.e. $\vec{n}=(n,0,\ldots,0)$, 
we find a {\em longitudinal length} $L_{\|}$. 
If $n$ is measured along one of the infinite directions, we find a {\em transverse length} $L_{\perp}(t)$.

The most simple example of a transverse length arises if the distances are measured along one of the coordinate axes in one of the infinite directions
(i.e. $\vec{n}=(0,0,\ldots,n)$ with $d^*\leq d-1$)
\BEQ
L_{\perp}^2(t) \:=\: \frac{ \sum_{n=-\infty}^{\infty} n^2 \exp\left[ -\pi \frac{n^2}{8\pi D t}\right]}{ 
\sum_{n=-\infty}^{\infty} \exp\left[ -\pi \frac{n^2}{8\pi D t}\right]}
\:\simeq\: 8\pi D t\: \frac{\int_{-\infty}^{\infty} \!\D n\: n^2\, e^{-\pi n^2}}{\int_{-\infty}^{\infty} \!\D n\:  e^{-\pi n^2}} \:=\: 4D\, t
\EEQ
which is identical to the known result for the bulk system \cite{Ebbi08}. 

A longitudinal length is found when $\vec{n}=(n,0,\ldots,0)$ with $d^*\geq 1$ is measured along one of the coordinate axes in a finite direction. 
If $N=2M$ is even, we have
\BEQ \label{A.29}
L_{\|}^2(t) 
=\frac{\sum_{n=-M+1}^{M} n^2\, \vartheta_3\bigl(\pi \frac{n}{2M},e^{-\pi/Z}\bigr)}{\sum_{n=-M+1}^{M}  \vartheta_3\bigl(\pi \frac{n}{2M},e^{-\pi/Z}\bigr)}
\EEQ
Using the definition of the Jacobi Theta function $\vartheta_3$, we have
\BEA
\lefteqn{\sum_{n=-M+1}^{M} \vartheta_3\bigl(\pi \frac{n}{2M},e^{-\pi/Z}\bigr) 
= \sum_{p\in\mathbb{Z}} \sum_{n=-M+1}^M \exp\left[ -\pi\II \frac{n}{M} p - \frac{\pi p^2}{Z} \right]}
\nonumber \\
&=& 2M + \sum_{p\ne 0} e^{-\pi p^2/Z} \left( 1 + e^{-\pi\II p} + \sum_{n=1}^{M-1} e^{-\pi \II (n/M) p} + \sum_{n=1}^{M-1} e^{\pi \II (n/M) p} \right) 
\:=\: 2M 
\label{A.30}
\EEA
and
\BEA
\lefteqn{\sum_{n=-M+1}^{M} n^2\, \vartheta_3\bigl(\pi \frac{n}{2M},e^{-\pi/Z}\bigr) 
\:=\: \sum_{p\in\mathbb{Z}} \sum_{n=-M+1}^M n^2\, \exp\left[ -\pi\II \frac{n}{M} p - \frac{\pi p^2}{Z} \right]}
\nonumber \\
&=& \sum_{n=-M+1}^M n^2 + \sum_{p\ne 0} e^{-\pi p^2/Z} \left( 0 + M^2 e^{-\pi\II p} + \sum_{n=1}^{M-1} n^2\, e^{-\pi \II (n/M) p} 
+ \sum_{n=1}^{M-1} n^2\, e^{\pi \II (n/M) p} \right) 
\nonumber \\
&\simeq& \frac{2}{3} M^3 + M^2 + \sum_{p\ne 0} e^{-\pi p^2/Z} \left(  M^2 (-1)^p  + \frac{4 (-1)^p}{\pi^2 p^2} M^3 + (-1)^p M^2 \right)  + {\rm O}(M) 
\nonumber \\
&\simeq& \frac{2}{3} M^3 + \frac{8 M^3}{\pi^2} \sum_{p=1}^{\infty} e^{-\pi p^2/Z} \frac{(-1)^p}{p^2} +{\rm O}\bigl(M^2\bigr) 
\label{A.31}
\EEA
where in the third line, an asymptotic expansion for $M$ large was made. 
Inserting (\ref{A.30},\ref{A.31}) into (\ref{A.29}) and fixing $8\pi D=1$ gives (\ref{5.9}). 
The same leading result also holds if $N=2M+1$ is odd.


\noindent
{\bf Acknowledgements:} It is a pleasure to thank H. Christiansen, W. Janke and S. Majumder for interesting discussions. I
also thank the MPIPKS Dresden (Germany) for warm hospitality, where this work was conceived. This work was supported by the
french ANR-PRME UNIOPEN (ANR-22-CE30-0004-01).  


\end{document}